# Evaluating the Financial Factors Influencing Maternal, Newborn, and Child Health in Africa

Er-Rays Youssef [1* [0000-0002-6691-9226]], M'dioud Meriem [2**[0000-0002-7855-3495]]

[1]Ibn Tofail University, Economics and Management Faculty (FEG), National School of Business and Management (ENCG), Research Laboratory in Organizational Management Sciences, Kenitra, Morocco
[2] Laboratory Engineering Sciences ENSA, Ibn Tofail University, Kenitra, Morocco
*raysyoussef@gmail.com    (CA)
** meriemmdioud@gmail.com

**Abstract:**

The study investigated the impact of healthcare system efficiency on the delivery of maternal, newborn, and child services in Africa. Data Envelopment Analysis and Tobit regression were employed to assess the efficiency of 46 healthcare systems across the continent, utilizing the Variable Returns to Scale model with Input orientation to evaluate technical efficiency. The Tobit method was utilized to explore factors contributing to inefficiency, with inputs variables including hospital, physician, and paramedical staff, and outputs variables encompassing maternal, newborn, and child admissions, cesarean interventions, functional competency, and hospitalization days.

Results revealed that only 26% of countries exhibited efficiency, highlighting a significant proportion of 74% with inefficiencies. Financial determinants such as current health expenditures, comprehensive coverage index, and current health expenditure per capita were found to have a negative impact on the efficiency of maternal-child services. These findings underscore a marginal deficiency in technical efficiency within Africa's healthcare systems, emphasizing the necessity for policymakers to reassess the roles of both human resources and financial dimensions in enhancing healthcare system performance.

**Keywords:** Healthcare systems; Africa; maternal, newborn, and child health; SDGs 2030; Data envelopment analysis; Tobit regression.

## 1    Background

Maternal, newborn, and child health (MNCH) is a crucial aspect of global well-being, as outlined in Sustainable Development Goals (SDGs) 3.1 and 3.2. However, Africa has the highest MNCH mortality ratio, with 287,000 women dying in 2020 and 5 million children dying in 2021. The World Health Organization (WHO) has identified regional imbalances in Africa, particularly in MNCH (Africa 2023). Global healthcare system evaluations are essential for identifying improvement areas, and recent research has focused on healthcare performance (Konca et Top 2023; WHO 2000). In African countries, MNCH faces significant pandemics due to weak leadership, corruption, and health system weaknesses. Despite the global mortality halving by 2021, profound inequalities persist (Murray et Frenk 2000; WHO 2021a).

DEA models are widely used to evaluate healthcare efficiency, as they evaluate the effectiveness of decision-making units (DMUs) that generate multiple outputs using multiple inputs(Chen et al. 2022; Yitbarek et al. 2019). Several studies have been conducted to compare health-care systems in various regions, including the 34 OECD member nations (Cetin et Bahce 2016), 30 European states (Asandului, Roman, et Fatulescu 2014), 20 Arab countries (El Husseiny 2022), the Middle East and North Africa (MENA) region (Hamidi et Akinci 2016), 18 nations within the MENA region (Meddeb 2019), and 46 Asian countries (Ahmed et al. 2019). In these studies, commonalities emerge as authors utilize variables such as health spending, doctors, and hospital beds. The outcomes of interest primarily focus on life expectancy and infant mortality rates. These consistent approach's, despite the diversity of regions and methodologies employed, significantly contributes to a comprehensive understanding of healthcare system efficiency on a global scale.

According to Kohl et al., whose discovered a relative drop in DEA research on the continent (Kohl et al., 2019), research on health system efficiency in Africa has been limited(Kohl et al. 2019). This approach was more commonly employed in African studies in the recent decade.

In 2023, Musoke et al. compared the health systems of twenty-nine least developed African countries. The inputs included domestic general government health, domestic private health, external health, and out-of-pocket health. outputs included the under-five survival rate, maternal survival ratio, life expectancy at birth, and infant survival rate (Musoke, Yawe, et Ssentamu 2023).

Top et al. examined 36 African healthcare systems, considering health expenditures in the GDP, medical professionals, nurses, and bed capacity per 1,000 individuals, the unemployment rate, and the Gini coefficient. Life expectancy at birth and 1/(infant mortality rate) were the study's output variables (Top, Konca, et Sapaz 2020).

Novignon and Novignon's study assessed the effectiveness of healthcare systems in 45 African countries using infant mortality rates and per capita health expenditure and real GDP(Novignon et Nonvignon 2017). Other study found that healthcare infrastructure in Sub-Saharan African countries is ineffective due to management weaknesses at multiple levels(Ibrahim et al. 2019). Kirigia's research investigated efficiency using factors like per capita total health expenditure, adult literacy rate, and male and female life expectancies as outcome variable(Kirigia 2015) and (Kirigia et al. 2007).

In a separate study, Arhin et al. assessed the ability of the health system to achieve the universal health coverage (UHC) goal by drawing evidence from 30 African countries. The study integrated per capita health spending, physician, and hospital data as inputs, with the UHC Index serving as the output metric (Arhin, Oteng-Abayie, et Novignon 2023).

However, Qu et al. undertook a comparative analysis encompassing 49 African countries from 2000 to 2017. They introduced an innovative methodology that amalgamates Data Envelopment Analysis (DEA) with the Gini coefficient to assess the efficacy of technology inequality in addressing environmental issues (Qu, Li, et N'Drin 2023).

The literature reviews an assessment of healthcare system efficiency in other regions, highlighting the need for a careful selection of inputs, outputs, and explanatory variables. Most of the studies used inputs, which included healthcare expenditures, healthcare personnel (doctors, nurses, midwives), hospital beds, and health facilities. These frequently employed outputs consist of life expectancy, healthcare utilization, and health outcomes. The most utilized explanatory variables include financial factors, governance, geographic location, infrastructure, and technology. However, most of these studies neglected the maternal mortality rate, stillbirth rate, neonatal mortality rate, and number of births attended by skilled health personnel. Hence, this original paper addresses the technical efficiency of the MNCH in Africa.

Motivated by the imperative to achieve SDGs 3.1 and 3.2 by 2030, it is paramount to assess the effectiveness of health systems in Africa, emphasizing the critical need for Africans to strengthen health system resilience. This research contributes significantly by providing insights into adopting best practices from more productive health systems, enriching the knowledge on productivity in resource-constrained settings, and presenting valuable literature for future researchers. The paper's originality lies in the meticulous selection of optimal and explanatory combinations, facilitating an assessment of the technical efficiency of forty-six healthcare systems in Africa, data envelopment analysis (DEA) and Tobit regression. The subsequent sections detail the structured literature review, methodology, results, discussion, conclusions, recommendations, limitations and future research.

## 2   Material and methods

### 2.1. Data sources and variables

This study included the latest data from the Global Health Observatory and World Health Organization (WHO) for 46 African countries, between 2005-2021 (WHO 2021b).

The input, output, and explanatory variables were selected to assess the accuracy of the WHO (WHO 2021b) statistics in reflecting the efficiency of the MNCH. Five inputs and outputs are considered to estimate technical efficiency (TE), which presented in table 1.

### 2.2. First stage: DEA

Technical efficiency (TE) is typically measured using two methods: parametric and nonparametric (Asmare et Begashaw 2018). A stochastic frontier production function based on a collection of explanatory variables is employed in the parametric approach. The nonparametric technique, on the other hand, employs linear programming to assess the relative efficiency of decision-making units (DMUs) by generating an ideal mix of inputs and outputs based on the best-performing unit in the collection (Asmare et Begashaw 2018) (Hollingsworth 2003).

Farrel introduced the DEA method (Farrell 1957), and Charnes et al. (1978) (Charnes, Cooper, et Rhodes 1978) and Banker et al. (1984) (Banker, Charnes, et Cooper 1984) developed this method. The most common technique is Data Envelopment Analysis (DEA), which may use it independently or in conjunction with a secondary analysis involving the Malmquist index (Malmquist 1953), Tobit regression (Tobin 1958a), and correlation efficiency (Babalola et Moodley 2020). Traditionally, two models are used to calculate the DEA: the CCR model developed by Charnes, Cooper, and Rhodes based on the assumption of constant returns to scale (Charnes et al. 1978) and the BCC model proposed by Charnes and Cooper based on the assumption of variable returns to scale (Banker et al. 1984). In the CRS model, outputs are assumed to increase proportionally with inputs, meaning that there are no economies or diseconomies of scale. This simplifies comparisons between similar-sized decision-making units (DMUs) (Banker et al. 1984; Charnes et al. 1978). In contrast, the VRS model allows for economies and diseconomies of scale, recognizing that each DMU may have an optimal operating size. This model is better suited for comparing DMUs of different sizes (Banker et al. 1984), as it isolates pure technical efficiency from the influence of scale DEA models can be categorized as either input-oriented or output-oriented, depending on the relationship between inputs and outputs.

DEA is a widely used method for assessing the relative efficiency of DMUs (Asmare et Begashaw 2018; Banker et al. 1984; Charnes et al. 1978; Farrell 1957; Hollingsworth 2003; Kuosmanen, Johnson, et Saastamoinen 2015; Malmquist 1953). It is particularly useful when there are multiple inputs and outputs involved in the evaluation process. DEA provides framework DMUs and those that achieve the highest level of output given a set of inputs (Chen et al. 2022; Yitbarek et al. 2019). In this study, the CRS and VRS were oriented (Chern et Wan 2000; Sherman et Zhu 2006).

The following formula shows the input-oriented VRS model, with results obtained using DEAP v2.1 (Coelli 1996). used in the study: (Banker et al. 1984; Top et al. 2020)

$$Max \frac{\sum_{r=1}^{S} u_{rk} y_{rj} + u_0}{\sum_{i=1}^{m} v_{ik} x_{ik}}$$

Contraints; $Max \frac{\sum_{r=1}^{S} u_{rk} y_{rj} + u_0}{\sum_{i=1}^{m} v_{ik} x_{ik}} \leq 1 (j, 1, 2, 3, \ldots, n)$

$$v_{rk}, v_{ik} \geq \varepsilon > 0, (r = 1,2,\ldots s), (i = 1, 2,\ldots,), u_0 \in R$$

## 2.3. Second stage: Tobit model.

The Tobit model is a statistical model commonly used when dealing with censored data. In this model, the dependent variable (Y) is subject to censoring, meaning that some values are not directly observed but fall within a certain range (Amemiya 1984) (Tobin 1958a).

The Tobit regression technique was used in this investigation, and the Excel data was transferred to STATA 18 for additional analysis. The standard Tobit model is illustrated in the following equation: (Tobin 1958b; Top et al. 2020)

$$yi *= x'_i \beta + u_i \ (i = 1, \ldots, n)$$
$$u_i \begin{cases} yi *, if \ yi * > 0 \\ 0, if \ yi * \leq 0 \end{cases}$$
$$u_i \sim IIN \ (0, \sigma^{-2})$$

In the formula, y* is a latent random variable that is observed as y if it is positive and is otherwise observed as equal to zero and the parameter vector β ∈ R$^k$. The error u$_i$ is an independent normal with a mean of zero and precision of σ$^2$ > 0.

We integrate two models. In the first model, the explanatory variables used are HBP, MD, NM, CHE, and CHEC (see the second model); in addition to the variables used in model 1, we use the variables OOPC, PVACC, CCI, CHEC and EXHC in model 2.

## 3. Results

Table 1 displays descriptive statistics for the study variables. Mean values for the key variables include 12.1 for HBP, 3.5 for MD, 15.2 for NM as Inputs. 23.5 for NMN, 18.7 for SB, 58.43 for U-5M, 41.5 for IMBA as outputs. 75.6 for BASHP, 354.2 for MMLB, 5.7 for CHE, 134.8 for CHEC, 17.2 for EXHC, 35.3 for OOPC, 67.7 for PVACC, and 49.1 for CCI as explanatory variables (see abbreviation).

Table 1. Descriptive analysis of and variables

|      | Input | | | Output | | | | Explanatory | | | | | | | |
|------|-----|-----|-----|-----|-----|-----|-----|-----|------|-----|------|-----|-----|-----|-----|
|      | HBP | MD  | NM  | NMN | SB  | U-5M | IMBA | BASHP | MMLB | CHE | CHEC | EXHC | OOPC | PVACC | CCI |
| Mean | 12.1 | 3.5 | 15.2 | 23.5 | 18.7 | 58.43 | 41.5 | 75.6 | 354.2 | 5.7 | 134.8 | 17.2 | 35.3 | 67.7 | 49.1 |

Figure 1 illustrates the technical efficiency of countries, including both efficient and inefficient ones. According to the Variable Returns to Scale (VRS) hypothesis, 12 states were deemed efficient, representing 26% with a score of 1. However, the majority, accounting for 76% of countries, scored less than 1. Notably, Gambia (19) exhibited low efficiency with a score of 0.403.

Additionally, the average efficiency score (TE-VRS) across all countries is 0.849 for VRS, signifying that healthcare systems across the African continent must minimize their inputs by 15% under an input orientation. Moreover, as the analyzed nations exhibited comparable outputs, those identified as efficient utilized relatively fewer resources than their inefficient counterparts. Eritrea (DMU 14) emerged as the most frequently referenced efficient country, being mentioned 34 times. From this perspective, Eritrea shares similarities with the inefficient countries in the input and output variables considered in this study. Seychelles (39 times) and Sao Tome (37 times) were the next most referenced efficient countries (refer to Figure 2).

According to the World Bank Income Classification (Table 2), when examining countries, those classified as high-income had the highest efficiency, representing only 13.04% (n=6), followed by countries with a score of 0.86; these countries constitute 39.13% (n=18). In the lowest classification, African countries had a score of 0.810, representing 43.47% (n=22) of the sample.

Figure 1: Input-oriented VRS scores

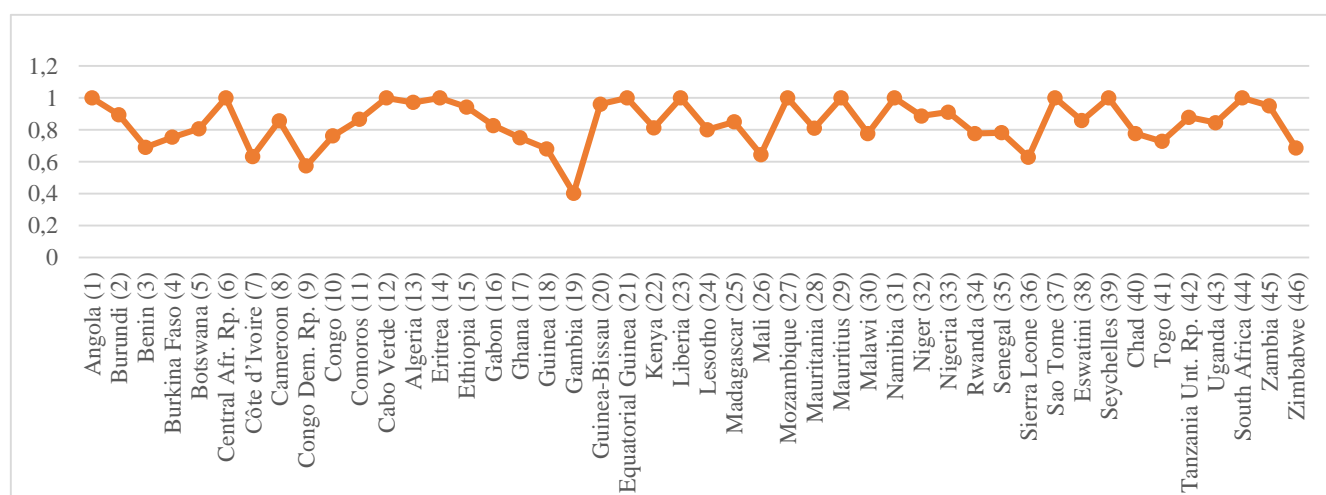

Table 2: World Bank Income Classification

| World Bank Income Classification | Average TE | DMU |
|---|---|---|
| Low income (LI)= 22 States | 0.810 | 2-4-6-9-11-14-15-18-19-20-23-25-26-30-32-34-36-40-41-43-45 |
| Low-middle income (LMI)= 18 States | 0.860 | 1-3-7-8-10-12-13-17-22-24-27-28-33-35-37-38-42-44-46 |
| High upper middle income (HUMI)b= 6 States | 0.940 | 5-16-21-29-31-39 |

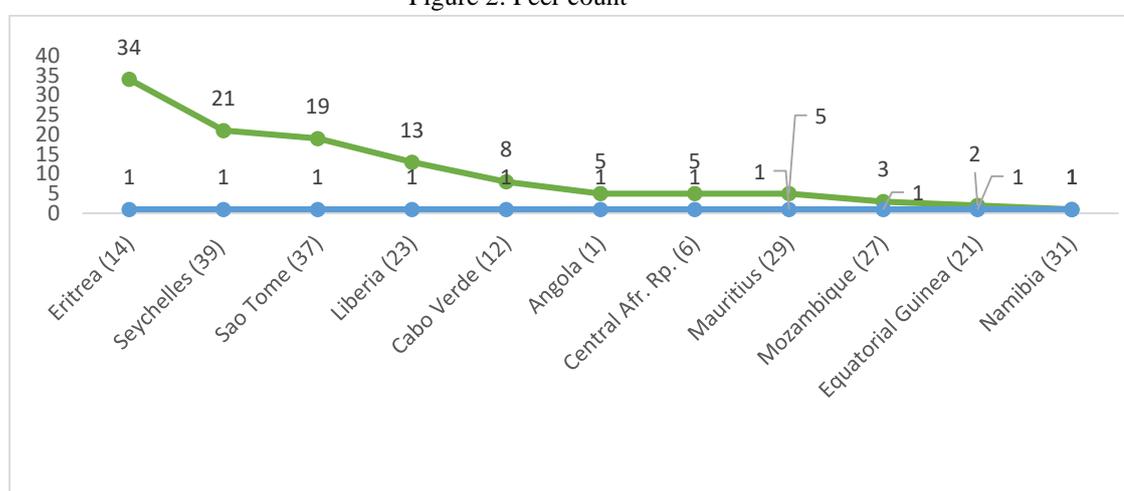

Figure 2: Peer count

The Tobit model was employed to investigate the underlying reasons for the reported inefficiency of healthcare systems in African states concerning maternal, newborn, and child health (MNCH). This model incorporated the explanatory variables listed in Table 3 and served as the second stage of analysis in this investigation. Two models were proposed.

At the 0.01 significance level, in both models, the healthcare expenditure (CHE) factor had a statistically significant effect on MNCH service inefficiencies, followed by the Comprehensive Country Index (CCI) variable in Model 2. Furthermore, at the 0.1 significance level, the combined effect of healthcare expenditure and corruption (CHEC) variable exhibited significance in Model 1.

**Table 3.** Tobit regression

|  | **Tobit Model 1** | **Tobit Model 2** |
|---|---|---|
| INF | Coef. Std. Err. t P>|t| | Coef. Std. Err. t P>|t| |

| | | |
|---|---|---|
| NM | .000891 .0044777 0.20 0.843 | .0014226 .0050997 0.28 0.782 |
| MD | -.006825 0191895 -0.36 0.724 | -.0123354 .0167856 -0.73 0.467 |
| HBP | -.004894 0084121 -0.58 0.564 | -.0027401 .0088742 -0.31 0.759 |
| CHE | -.063752 .0216531 -2.94 0.005* | -.0811233 .023045 -3.52 0.001* |
| CHEC | -.004037 .0023717 -1.70 0.096*** | -.0019306 .0020531 -0.94 0.353 |
| PVACC | | -.000039 .001852 -0.02 0.983 |
| CCI | | .0155259 .0041438 3.75 0.001* |
| OOPC | | -------------------------------------- |
| EXHC | | -.004892 .004223 -1.16 0.254 |
| cons | .7135877 .1277524 5.59 0.00 | .08196 .1909008 0.43 0.670 |
| sigma | .26837 .0332456 | 2264282 .0276046 |
| LR chi2(5) | 23.37 | 39.07 |
| Prob > chi2 | 0.0003 | 0 |
| Log likelihood | -10.158682 | -2.307881 |
| Pseudo R2 | 0.5349 | 0.8943 |

*: 0.01, **: 0.05, and ***: 0.1

## 3. Discussion and Conclusion

Health systems aim to ensure equitable public access to healthcare services and judicious resource distribution. The responsibility for funding these requirements lies with the public. The SDGs for 2030 urge governments to adopt reforms to enforce regulations in this realm, as emphasized by SDG 3. Most maternal, newborn, and child health (MNCH) services rely on health system resources, and the SDGs emphasize the need for efficient funding. This study analyzed the MNCH service efficiency of 46 countries in Africa in the context of the SDGs, utilizing the DEA method in the first stage and Tobit regression in the second stage.

The research findings disclose a disconcerting scenario, elucidating a substantial dissonance between the prevailing maternal and child health metrics in Africa and the specified SDGs for the year 2030. Specifically, the average maternal death rate in these countries was reported at 354.17 per 100,000 live births in 2021, whereas the targeted SDGs (3.1) for 2030 stand at 70 per 100,000 live births.

Moreover, in the domain of newborn and child health, the data revealed that the average under-five mortality rate (U-5M) was 58.43 per 1000 live births, a figure significantly higher than the SDGs target (3.2) of 25 per 1000 live births. Additionally, regarding neonatal mortality (NMN), the findings indicated an average rate of 23.5 per 1000 live births, a substantial disparity from the United Nations (SDG) goal of 12 per 1000 live births.

This alarming disparity between the observed metrics and the established SDG targets underscores the considerable distance that African countries currently find themselves from realizing the objectives outlined in SDG 3. Addressing this discrepancy necessitates a comprehensive evaluation of the efficiency and various influencing variables within the maternal, newborn, and child health (MNCH) domain.

The findings from the DEA analysis revealed notably low or medium efficiency for most African countries. This suggests that 22 out of 46 states represent low-income countries, followed by 18 out of 46 states classified as low-medium income. Eritrea was the most referenced country. According to the Tobit model analysis, financial factors such as healthcare expenditure (CHE), Comprehensive Country Index (CCI), and the combined effect of healthcare expenditure and corruption (CHEC) had negative effects on the inefficiency of the health system related to MNCH. This indicated that the health financing system suffers from profound dysfunctions, which hinder the promotion of MNCH in African countries.

According to previous studies on African countries, the performance of health systems was generally low or moderately efficient based on scores (Africa 2023; Ibrahim et al. 2019; Kirigia 2015; Musoke et al. 2023; Qu et al. 2023; Top et al. 2020). The World Health Organization (WHO) reported an average technical efficiency score of 0.79 across its 47 member countries in 2019 (Africa 2023). Ibrahim et al. assessed healthcare systems in Sub-Saharan Africa and identified them as inefficient overall. Over the analyzed period, only three provinces in 2015, Rwanda in 2014 and 2015, and Tanzania in 2015, were deemed efficient (Ibrahim et al. 2019).

The study also discovered that governance metrics, notably the rule of law and government efficacy, have a greater impact on healthcare system efficiency than public health spending. This implies that effective resource management is more important than the amount of money invested in healthcare systems in Sub-Saharan African nations (Ibrahim et al. 2019).. According to Babalola and Moodley's findings, less than 40% of the facilities tested were efficient. These studies reported parameters such as catchment population, facility ownership, and geography (Babalola et Moodley 2020).
Arhin et al. discovered that by implementing best practices in instruction, management performance, expenditures on public health, external health funding, and prepayment arrangements, 30 Sub-Saharan African health systems can increase Universal Health Coverage (UHC) levels by 19% while using existing healthcare resources (Arhin et al. 2023).

The overall healthcare efficiency in different African countries is considerable, notably Ghana, Sierra Leone, and Burkina Faso all recorded a low technical efficiency score in the provision of MNCH (Ibrahim et al. 2019; Marschall et Flessa 2011; Top et al. 2020). The choices of input and output variables depend on the availability of information in the reports concerning the activities of health establishments in these countries. Technical efficiency varies from one health system to another.
In Ghana's case, 78% of primary healthcare institutions have a low efficiency score (Akazili et al. 2008). Primary healthcare facilities in KwaZulu-Natal, South Africa, similarly have a 70% low technical efficiency (Kirigia, Sambo, et Scheel 2001). Al-Hassan et al. found that the geographical location of the centers and the type of ownership were substantially connected with the prediction of efficiency scores rather than the quality of service(Alhassan et al. 2015). Marschall et al.'s Tobit model results in Burkina Faso demonstrated that the explanatory variables determining inefficiency in rural healthcare were highly related to geographical distance and other factors (Marschall et Flessa 2011).

The management of African healthcare systems, particularly in the realm of maternal, newborn, and child health (MNCH), presents a multifaceted challenge encompassing economic, social, political, and infrastructural factors. These challenges include financial constraints, human resource shortages, infrastructure deficiencies, cultural and social barriers, governance issues, high disease burdens, inadequate health facility capacity, suboptimal utilization of health services, leakages, and corruption. Economically advanced countries such as Eritrea, Seychelles, Mauritius, Namibia, South Africa, and Sao Tome exhibit efficient health systems. However, economically less developed countries encounter difficulties in providing and accessing health services due to their developmental status and less robust institutional frameworks.

A literature review revealed that countries like Ghana, Sierra Leone, and Burkina Faso all demonstrated low-efficiency scores in delivering MNCH services. It is imperative to advocate for enhanced resource allocation strategies, prioritize efficient utilization of healthcare resources, optimize infrastructure enhancements, invest in workforce training, and embrace technology to streamline service delivery. Health authorities are urged to consider comprehensive policy reforms aimed at addressing operational inefficiencies identified in the study. These reforms should be strategic and tailored to enhancing the overall effectiveness of healthcare systems in the domain of maternal and child health.

The study assessed the effectiveness of healthcare systems; however, its precision relies on data from the World Health Organization (WHO), which may overlook key determinants influencing MNCH outcomes. Additionally, the study assumes homogeneity in production functions across diverse African countries, potentially oversimplifying variations in healthcare infrastructure, socioeconomic conditions, and cultural factors. Moreover, the study's focus on internal factors may neglect external influences such as political stability and global health crises. Generalizing the findings beyond the studied nations is also risky due to the continent's heterogeneity and the dynamic nature of efficiency. Future research could explore sustainable financing solutions for healthcare systems, addressing structural constraints faced by African states.

## Declaration

### Abbreviation

| Variable | Abbreviation | SDGs |
|---|---|---|
| HBP | Hospital beds (per 10 000 population) | 3.c.1 workforce |
| MD | Medical doctors (per 10,000) | |
| NM | Nursing and midwifery personnel | |
| CHEC | Current health expenditure (CHE) per capita in US$ 2020 | 3.c. Health financing |
| EXHC | External health expenditure (EXT) per capita in US$ 2021 | |
| NMN | Neonatal mortality rate (per 1000 live births) 2021 | 3.2. Neonatal and child mortality |
| SB | Stillbirth rate (per 1000 total births) 2021 | |
| IMBA | Infant mortality rate (probability of dying between birth and age 1 per 1000 live births) | |
| BASHP | Births attended by skilled health personnel (%) | |
| MMLB | Maternal mortality ratio (per 100 000 live births) 2020 | 3.1. Maternal mortality |
| PVACC | Proportion of vaccination cards seen (%) | |
| CCI | HeaB1: Reproductive, maternal, newborn and child health interventions (RMNCH), combined Composite coverage index (%) | |
| DEA | Data Enveloppement Analysis | |
| VRS | Variable Returns Scale | |
| TE | Technical Efficiency | |
| DEAP | Data Envelopment Analysis Programming | |
| DMU | Decision Making Unit | |
| SDGs | Sustainable Development Goals | |
| MNCH | maternal, newborn, and child health | |
| WHO | World Health Organization | |

## Statements and Declarations

### Abbreviations

CRS: Constant Returns Scale, DEA: Data Envelovment Analysis, DEAP: Data Envelovment Analysis Programming, DMU: Decision Making Unit, IM: Index Malmquist, Effch: Technical Efficiency Change, ET: Technical Efficiency, pech: Pure Change, MNCH: maternal, newborn and child Hospital, SE: Scale Efficiency, sech: Scale Change, Techch: Technology Change, TFP: Total Factor Production, tfpch: Total Factor Productivity, VRS: Variable Returns Scale.

### Ethical approval and consent to participate

Not applicable

### Competing interests

The authors declare that they have no competing interests.

**Authors' contributions**

The authors were involved in the literature review, data analysis, interpretation of the results, and drafting of the manuscript. The author read and approved the final manuscript.

**Acknowledgements**

We are immensely grateful to the Health Ministry Moroccans for their cooperation in collecting the data and making available the Annual Health Hospital Activity Reports 2017, 2018, 2019, and 2020. We are also thankful to Ibn Tofail University for facilitating the coordination of the data collection for the study.

**Availability of data and material**

Data from the Health Ministry Moroccan 2012, 2013, 2014, 2015, 2016, 2017, 2018, 2019 and 2020

**Competing interests**

The authors declare that they have no competing interests.

**Consent for publication**

Not applicable

**Funding**

Not applicable

**Author details**

Received:
Accepted:
Published Online: